**Artificial two-dimensional polar metal by charge transfer to a ferroelectric insulator**

W. X. Zhou[1,2,†], H. J. Wu[3,†], J. Zhou[1,2], S. W. Zeng[1,2], C. J. Li[1,3], M. S. Li[3], R. Guo[1,3], J. X. Xiao[1,3], Z. Huang[1,2], W. M. Lv[1], K. Han[1,2], P. Yang[4], C. G. Li[5], Z. S. Lim[1,2], H. Wang[1,3], Y. Zhang[3], S. J. Chua[5], K. Y. Zeng[6], T. Venkatesan[1,2,5,7], J. S. Chen[1,3], Y. P. Feng[2], S. J. Pennycook[3], A. Ariando[1,2,7]*

[1]NUSNNI-NanoCore, National University of Singapore, Singapore 117411.

[2]Department of Physics, National University of Singapore, Singapore 117542.

[3]Department of Materials Science and Engineering, National University of Singapore, Singapore 117575.

[4]Singapore Synchrotron Light Source (SSLS), National University of Singapore, 5 Research Link, Singapore 117603.

[5]Department of Electrical and Computer Engineering, National University of Singapore, Singapore 117576.

[6]Department of Mechanical Engineering, National University of Singapore, 9 Engineering Drive 1, Singapore 117576.

[7]NUS Graduate School for Integrative Sciences and Engineering, National University of Singapore, Singapore 117456.

[†]These authors contributed equally to this work

*To whom correspondence should be addressed.

Email: ariando@nus.edu.sg

**Integrating multiple properties in a single system is crucial for the continuous developments in electronic devices. However, some physical properties are mutually exclusive in nature. Here, we report the coexistence of two seemingly mutually exclusive properties-polarity and two-dimensional conductivity-in ferroelectric $Ba_{0.2}Sr_{0.8}TiO_3$ thin films at the $LaAlO_3/Ba_{0.2}Sr_{0.8}TiO_3$ interface at room temperature. The polarity of a ~3.2 nm $Ba_{0.2}Sr_{0.8}TiO_3$ thin film is preserved with a two-dimensional mobile carrier density of ~0.05 electron per unit cell. We show that the electronic reconstruction resulting from the competition between the built-in electric field of $LaAlO_3$ and the polarization of $Ba_{0.2}Sr_{0.8}TiO_3$ is responsible for this unusual two-dimensional conducting polar phase. The general concept of exploiting mutually exclusive properties at oxide interfaces via electronic reconstruction may be applicable to other strongly-correlated oxide interfaces, thus opening windows to new functional nanoscale materials for applications in novel nanoelectronics.**

Oxide interfaces provide a fertile ground for multifunctional integration because the delicate balance between spin, orbit, charge and lattice degrees of freedom in oxides can be easily destabilized with relatively small stimuli[1-4]. Ground-breaking examples are the coexistence of ferromagnetism and superconductivity at the LaAlO$_3$/SrTiO$_3$ (LAO/STO) interface although ferromagnetism is expected to destroy the pairing interaction responsible for superconductivity[5-6], and the coexistence of ferromagnetism and ferroelectricity at Fe/BaTiO$_3$ interface although ferroelectricity normally requires empty *d* orbitals while ferromagnetism is usually a result of partially filled *d* orbitals[3,7]. Recently, polar metals, which possess another pair of seemingly mutually exclusive properties, ferroelectric-like polarity and conductivity, has attracted a flurry of interest. Traditionally, it was considered that introducing itinerant electrons into a ferroelectric can eliminate ferroelectricity and associated polarity as the electrons screen the long-range Coulomb forces, which favor a polar structure[8-11]. However, recent first-principles calculations show that the charge rearrangements associated to electrostatic screening induces local lattice response, which favors polar distortions and that local off-centering can be sustained up to ~0.1 electron per unit cell (e/uc)[11-13], suggesting that there is no fundamental incompatibility between metallicity and polar distortions.

In recent decades, the research on polar metals has been mainly on three routes: (i) Searching for native polar metals, such as LiOsO$_3$[14], Bi$_5$Ti$_5$O$_{17}$[15], Cd$_2$Re$_2$O$_7$[16] and WTe$_2$[17]; (ii) Doping charges into ferroelectric insulators, such as BaTiO$_{3-\delta}$[10,11] and CaTiO$_{3-\delta}$[12]; and (iii) Stabilizing a polar phase in an otherwise non-polar metal (NdNiO$_3$) by deliberate geometric design[18]. However, most of the research was focused on bulk polar metals and the question of whether such coexistence can occur in a two-dimensional system remains largely unexplored. In this report, we directly show that polarity and two-dimensional electron gas (2DEG) can coexist in a single phase by charge transfer doping to a ferroelectric insulator.

A promising candidate to study this problem is the interface between a "polar" oxide (LAO) and a ferroelectric ($Ba_{0.2}Sr_{0.8}TiO_3$). In essence, ferroelectricity arises from the charge separation between positive and negative ions, which create a spontaneous polarization and an internal electric field. Similar to the charge separation in ferroelectrics, in the "polar" oxide LAO, the physically separated alternating stacking of charge-positive $(LaO)^+$ and charge-negative $(AlO_2)^-$ layers also creates a built-in electric field. The possible charge transfer generated by the potential across $LaAlO_3/Ba_{0.2}Sr_{0.8}TiO_3$ (LAO/BST) interface created by the difference between the electric fields in LAO and BST can thus serve as a useful method to dope electrons into ferroelectric BST. This charge transfer is possible because the polar-discontinuity, which generates the 2DEG at the LAO/STO interface is also present at the LAO/BST interface as BST have a stacking of charge-neutral $Ti^{4+}O_2^{2-}$ and $Sr^{2+}O^{2-}$ (or $Ba^{2+}O^{2-}$) layers similar to that of STO[19,20]. We use ferroelectric BST thin films instead of the conventional $BaTiO_3$ (BTO) used in previous studies[10,11] as we found that the Coulomb forces in BTO strongly localize electrons (see supplementary note 1 for more details), while doping Sr in BTO is known to weaken the ferroelectricity and hence improving electron mobility.

**Results**

**Ferroelectricity of BST thin films on conducting Nb-doped STO substrate.** We first check the ferroelectricity of the BST thin films by performing hysteresis loops, domain writing and reading experiments on a 10 uc BST/Nb:STO (001) sample using piezoresponse force microscopy at room temperature (Fig. 1a). In its bulk form, the Curie temperature of BST is 105 K[21]. However, we find that the as-grown BST film is ferroelectric at room temperature in a single-domain state with a downward spontaneous polarization (pointing from BST to Nb:STO). This large Curie temperature enhancement is probably a result of strain[22] and dimensional effects[23]. The 180° phase change between 4 and -7 V (Fig. 1b,c) and the good domain writing and reading with both positive and negative voltages (Fig. 1d,e) indicate robust ferroelectricity in the BST thin film.

The -1.5 V voltage shift in the phase and amplitude hysteresis loops is due to the well-known imprint effect as commonly observed in other ferroelectric thin films[24-26]. This could be collectively caused by the following mechanisms: (i) Strain effect imposed on BST thin film by Nb:STO substrate[24], (ii) Asymmetric electrostatic boundary condition caused by the different work functions between the two electrodes[25], (iii) Defect dipole accumulation at the BST/Nb:STO interface due to interfacial diffusion of chemical species[26]. Note that in Fig. 1e, some partial relaxation is observed, which is attributed to polarization relaxation, as is commonly observed in other ferroelectric ultrathin films, such as $BaTiO_3$[27].

**Ferroelectricity modulated electronic transport properties of the 2DEG.** Having observed room temperature ferroelectricity in BST thin film, we grew BST and LAO successively on atomically flat $TiO_2$-terminated STO (001) substrates by pulsed laser deposition to form a LAO/BST/STO heterostructure (see supplementary notes 2-4 for more details), where a 2DEG is observed at the LAO/BST interface. On one hand, the itinerant electrons introduced by the polar discontinuity at the LAO/BST interface tend to destroy the ferroelectricity and associated polarity of BST. On the other hand, the downward polarization of BST creates a downward electric field in LAO, which obstructs the polar-discontinuity-induced charge transfer from the LAO surface to the LAO/BST interface. The competition between the built-in electric field of LAO and the polarization of BST is exploited by varying the magnitude of the polarization of BST with different thicknesses. Figure 2 shows the electronic transport properties of a set of samples with fixed LAO thickness (15 uc) and different BST thicknesses. Figure 2a shows temperature dependent sheet resistance ($R_s$-$T$) for these metallic samples. It should be noted that a single layer of BST with varying thicknesses deposited on STO substrate is insulating, indicating that a LAO layer is required to obtain conductivity in LAO/BST/STO heterostructure. Above 100 K, sheet resistance gradually increases with increasing BST thickness. However, a crossover in the $R_s$-$T$ curves can be observed around 20-50 K and the sheet resistance dependence on BST thickness is reversed at low temperatures (2-20 K).

Carrier density gradually decreases with increasing BST thickness as the magnitude of the polarization increases (Fig. 2b)[27]. Another feature in Fig. 2b is that the temperature dependent carrier density ($n$-$T$) curves become less temperature dependent with increasing BST thickness and finally turns into (nearly) temperature independent when BST exceeds 8 uc (see supplementary notes 5-6 for more details). The decreasing carrier density with decreasing temperature in STO is attributed to freeze-out of oxygen-vacancy-induced carriers[20]. The decreasing carrier density dependence on temperature indicates that the oxygen vacancies in our samples decrease with increasing BST thickness. This is confirmed by the decreasing photoluminescence (PL) intensity with increasing BST thickness in our PL measurements (see supplementary note 7 for more details), where the PL intensity is proportional to the oxygen vacancy concentration in STO. In addition, theoretical calculations show that under reducing conditions, the oxygen vacancy formation energy in ferroelectric BTO is higher than that in STO[28]. Hence, we conclude that the oxygen vacancy formation energy in ferroelectric BST is higher than that in paraelectric STO. When the BST thickness is below 8 uc, a 2DEG resides in both STO and BST, and a large number of oxygen-vacancy-induced carriers are present in STO. In this case, the decreasing carrier density with increasing BST thickness is a result of both the electric field effect and reduced oxygen vacancy concentration. When the BST thickness exceeds 8 uc, the 2DEG lies only in the BST and few oxygen-vacancy-induced carriers are present, leading to observation of temperature-independent carrier density. In this case, the transport properties are influenced only by the competition between the built-in electric field of LAO and the polarization of BST without the influence of oxygen vacancies, signifying the coexistence of polarity and the 2DEG above 8 uc of BST. The carrier density of the 8 uc sample is ~$3\times10^{13}$ cm$^{-2}$, which corresponds to ~0.05 e/uc. Here, we estimate the thickness of the 2DEG to be 8 uc (~3.2 nm). This is in good agreement with previous observations, which showed that the thickness of the 2DEG at the LAO/STO interface is around 2-4 nm[29,30]. As discussed above, both carrier density and oxygen vacancies decrease with increasing BST thickness, leading to less electron-electron scattering and defect scattering, which consequently account for the mobility enhancement with increasing BST thickness (Fig. 2c,d).

We note the results in Fig. 2 are strikingly similar to the electric field gating experiments of LAO/STO interface[31,32], suggesting that the 2DEG is modulated by the electric field provided by the ferroelectric BST. We also note that the upturn of the $R_s$-$T$ curve and the corresponding decreasing carrier density with decreasing temperature of the 15 uc BST sample suggest carrier localization and is consistent with Mott variable-range hopping model (see supplementary note 8 for more details). This behavior could be due to the strong expulsion of electrons as the BST becomes thicker, as well as increasing interfacial disorder with increasing film thickness and strain effects introduced by substrate misfit as observed in epitaxial LAO/STO systems[33].

**Direct observation of the coexistence of polar displacements and 2DEG.** We next turn to characterize the structural, elemental and electronic features of a 15 uc LAO/10 uc BST/STO heterostructure with scanning transmission electron microscopy (STEM), coupled with electron energy loss spectroscopy (EELS). We detected the positions of A-site (A = La, Ba or Sr), B-site (B = Al or Ti) and oxygen atoms from the contrast-inverted annular-bright-field (ABF) STEM image (Fig. 3a,b and supplementary note 9). The BST layer shows a downward polarization, which is consistent with the as-grown BST thin film on Nb:STO (Fig. 1). The B-O$_{II}$ and A-O$_I$ displacements are largest at the LAO/BST interface with magnitudes as high as ~ 40 and 25 pm, respectively, which are comparable to the displacements of PbZr$_{0.2}$Ti$_{0.8}$O$_3$ thin films on STO substrate[34]. The displacements gradually decrease into both LAO and STO and becomes negligibly small far away from the LAO/BST interface (Fig. 3c). The remnant displacement in LAO is probably due to BST diffusion into LAO (Fig. 4) and/or the buckling effect commonly observed in LAO/STO heterostructures[35-37].

STEM-EELS spectrum imaging (Fig. 3d-g) was employed to reveal the layer-by-layer elemental distribution and electronic structure by tracing the Ti $L_{2,3}$ (Fig. 3e) and O K (Fig. 3f) edges from the STO substrate to the top LAO layer. The BST/STO interface is atomically sharp, while some Ti diffusion (~ 1 uc) into the LAO layer at the LAO/BST interface is observed (Fig. 3d). The Ti valence evolution from STO and BST to the LAO/BST interface is clearly observed from the red-shift of the peak around 465 eV (Fig. 3e). In STO and BST regions far away from the LAO/BST interface, each Ti $L_{2,3}$ curve is a typical $Ti^{4+}$ spectrum, while around the LAO/BST interface, each Ti $L_{2,3}$ curve is a mixture of $Ti^{4+}$ and $Ti^{3+}$ spectra. The layer-by-layer relative ratio $Ti^{3+}/(Ti^{3+} + Ti^{4+})$ grows gradually from BST to the LAO/BST interface and reaches ~ 0.2 at the interface (Fig. 3g), as evidenced via multiple linear least squares (MLLS) fitting of the experimental spectra (see supplementary note 10 for more details)[38]. Another way to trace the valence state of Ti is from the O K edge, which also shows significant changes when the Ti oxidation state changes across the interface (Fig. 3f)[37]. The energy difference between peak A and C ($\Delta E$) has been recognized as an accurate indicator of valence change in perovskites[37], the decrease of $\Delta E$ from BST to LAO/BST interface is evident in Fig. 3f-g and is related to a decrease in the Ti valence. These results unambiguously reflect a decrease of Ti valence from $Ti^{4+}$ in BST to $Ti^{3+}$ at the LAO/BST interface, indicating that the 2DEG lies at the LAO/BST interface and the carrier density of the 2DEG decreases away from the LAO/BST interface into the BST (Fig. 3g). The STEM and EELS results directly show that the polar displacements and $Ti^{3+}$ (*i.e.* excess electrons) coexist at the LAO/BST interface in a single phase without any detectable phase separation.

**First-principles calculations of the 2DEG and polar displacements in LaAlO$_3$/Ba$_{0.5}$Sr$_{0.5}$TiO$_3$/SrTiO$_3$ heterostructure.** In order to investigate the relation between the charge density and polarity in LAO/BST/STO heterostructure, we performed density-functional theory (DFT) based calculations. The profile of the conducting electron density and the displacements between anions and cations for the whole LAO/BST/STO heterostructure are shown in Fig. 4. For simplicity, instead of using a LaAlO$_3$/Ba$_{0.2}$Sr$_{0.8}$TiO$_3$/SrTiO$_3$ heterostructure as in the experiments, we used a LaAlO$_3$/Ba$_{0.5}$Sr$_{0.5}$TiO$_3$/SrTiO$_3$ heterostructure.

Our first-principles calculations were performed using density-functional theory based Vienna ab initio Simulation Package (VASP)[39,40] with the local density approximation (LDA) for the exchange-correlation functional[41] and the frozen-core all-electron projector-augmented wave (PAW) method[42,43]. The cutoff energy for the plane wave expansion is set to 400 eV. Monkhorst-Pack k-point grids for Brillouin zone sampling are set to 6×6×6 for bulk SrTiO$_3$, 6×6×6 for bulk LaAlO$_3$, 5×5×6 for bulk Ba$_{0.5}$Sr$_{0.5}$TiO$_3$, and 5×5×1 (10×10×1) for LaAlO$_3$/Ba$_{0.5}$Sr$_{0.5}$TiO$_3$/SrTiO$_3$ superlattice structure relaxation (static calculations), respectively. The structures are fully relaxed until the forces are smaller than 0.01 eVÅ$^{-1}$ for the bulk LaAlO$_3$, Ba$_{0.5}$Sr$_{0.5}$TiO$_3$, SrTiO$_3$, and 0.02 eVÅ$^{-1}$ for LaAlO$_3$/Ba$_{0.5}$Sr$_{0.5}$TiO$_3$/SrTiO$_3$ superlattice structure. The equilibrium lattice constant for SrTiO$_3$ is 3.865 Å. We fix the in-plane lattice constant of LaAlO$_3$ and Ba$_{0.5}$Sr$_{0.5}$TiO$_3$ to this number and optimize their $c$ axis into 3.66 Å and 3.96 Å respectively. For Ba$_{0.5}$Sr$_{0.5}$TiO$_3$, the ratio of Ba and Sr is set to 1:1. We use $\sqrt{2} \times \sqrt{2} \times 1$ supercell for LaAlO$_3$/Ba$_{0.5}$Sr$_{0.5}$TiO$_3$/SrTiO$_3$ superlattice, which is a symmetric structure with 4.5 uc SrTiO$_3$ in the middle, 5 uc Ba$_{0.5}$Sr$_{0.5}$TiO$_3$ on each of its two TiO$_2$ terminals, and then 3.5 LaAlO$_3$ on top of Ba$_{0.5}$Sr$_{0.5}$TiO$_3$. The charge density is calculated by performing a partial charge calculation of the conduction bands below Fermi level.

The electrons mainly locate at Ti atoms in both Ba$_{0.5}$Sr$_{0.5}$TiO$_3$ and SrTiO$_3$ with the largest charge density at LaAlO$_3$/Ba$_{0.5}$Sr$_{0.5}$TiO$_3$ interface (0.21 e/uc), and decays gradually from this interface into SrTiO$_3$ (Fig. 4a). This result agrees both qualitatively and quantitatively with the Ti$^{3+}$ fraction analysis in Fig. 3f. The B-O$_{II}$ and A-O$_I$ displacements, which are in downward polarization (see supplementary note 11 for more details), are largest at the LaAlO$_3$/Ba$_{0.5}$Sr$_{0.5}$TiO$_3$ interface and gradually decreases into LaAlO$_3$ and SrTiO$_3$ (Fig 4b). The trend of the calculated layer-dependent displacements well reproduces the STEM results in Fig. 3b. We note that the absolute values of the calculated displacements are around one half of those from experimental observations. This may be attributed to the underestimation of lattice constant and polarization by LDA functional, possible defects in the samples and/or measurement errors in STEM-extracted displacements. Figure 4c shows the structural guide of the supercell used in the calculations.

**Discussion**

Several mechanisms may contribute to the behaviour of the layer-dependent displacements observed in Fig. 3 and 4: (a) A depolarizing buckling, which has been widely reported in LAO/STO heterostructure, could occur in BST to compensate the built-in electric field in LAO[35-37]. (b) Screening of the depolarization field of BST by the insulating SrTiO$_3$ substrate. In an earlier report, Chisholm et al. reported that the depolarization field of PbZr$_{0.2}$Ti$_{0.8}$O$_3$ in PbZr$_{0.2}$Ti$_{0.8}$O$_3$/SrTiO$_3$ heterostructure could be screened by oxygen vacancies in SrTiO$_3$[34]. This screening mechanism could also lead to diminishing displacement in the first few layers of BST away from BST/STO interface. (c) Screening of ferroelectricity by itinerant electrons. As proposed by Wang et al., the critical electron density for stable polar displacement in BaTiO$_3$ is 0.11 e/uc[11]. Here, we observed large Ti-O displacement (0.17 Å) at the LaAlO$_3$/ Ba$_{0.5}$Sr$_{0.5}$TiO$_3$ interface with an electron density ~0.21 e/uc in Fig. 4. The discrepancy between our results and those reported by Wang et al. may be attributed to the substrate-imposed compressive strain, which has been reported to be a useful way to increase the critical electron density in BTO[44]. We also note that the carrier density calculated from DFT and that extracted from EELS measurement are larger than the experimental observations (0.05 e/uc). This is probably because that some of the charges are localized in the first few layers of BST at LAO/BST interface as commonly observed at LAO/STO interface[29,30]. To disentangle these inter-correlated mechanisms, more detailed experimental and theoretical work needs to be done. Nevertheless, the overall effect in Fig. 3 and 4 clearly suggests the coexistence of 2DEG and polarity in a single phase.

Our discoveries demonstrate a route to create a two-dimensional polar metal at oxide interface through interfacial electronic reconstruction, which is achieved by deliberately engineering the competition of the electric fields between a ferroelectric and a "polar" oxide. As coexistence of ferromagnetism and superconductivity has already been demonstrated at the LAO/STO interface[5,6], the integration of polarity further expands the functionality of this interface and offers new opportunities for future multifunctional devices. Moreover, the ferroelectric soft phonons could be utilized to stabilize the superconducting phase at elevated temperatures[45]. We notice that during the review process of our work, a similar work by Cao et al. reported the coexistence of polarity and 2DEG in a tri-color $BaTiO_3/SrTiO_3/LaTiO_3$ heterostructure and we highly recommend this work to the readers of our work[46]. Finally, we note that a recent publication reported an electrically switchable ferroelectric topological semimetal $WTe_2$[17]. Nevertheless, the switchability remains elusive in doped complex-oxide-based ferroelectrics, despite a previous theoretical report proposed a promising candidate $Bi_5Ti_5O_{17}$[15].

**Methods**

**Sample preparation.** The samples were grown by pulsed laser deposition (PLD) equipped with *in-situ* reflection high-energy electron diffraction (RHEED). BST layer and LAO layer were successively deposited on atomically flat $TiO_2$-terminated STO (001) substrates at 760 °C and an oxygen pressure of $10^{-4}$ torr. During the deposition, a KrF laser with a wavelength of 248 nm was used. The laser repetition was 1 Hz and the laser energy density was 1.5

J/cm$^2$. The 10 uc BST/Nb:STO (0.1 wt% Nb doped) sample was prepared under the same conditions except that no LAO was deposited. To obtain TiO$_2$-terminated surface, STO and Nb:STO substrates were treated with buffered hydrofluoric acid for 30 seconds followed by annealing at 950 °C for 1.5 h in air[47].

**Piezoresponse force microscopy.** Ferroelectric properties were measured using piezoresponse force microscope (PFM) and piezoresponse force spectroscopy (PFS) techniques. In this study, a commercial SPM system (MFP-3D, Asylum Research 13.03.70, USA) was used, which was controlled by a commercial software (IGOR PRO 6.34A). In the PFM measurements, amplitude ($A$) and phase ($\phi$) images can be obtained simultaneously by applying an AC bias to the tip. The applied AC bias was 0.2 V. The amplitude image refers to piezoresponse of the material, whereas the phase image regards to the polarization direction of the material. Before the PFM measurements, DC writing processes were conducted. In these processes, dc biases of -8/8 V were sequentially applied to the conductive tip. The corresponding written areas were 3×3 and 1×1 µm$^2$, respectively. In the Piezoresponse Force Spectroscopy (PFS) measurement, the PFM was operated in the spectroscopy mode, the tip with a triangle-square waveform was fixed at a certain location. The frequency of the used waveform was 200 mHz, i.e., the bias-on and bias-off time was 25 ms. To exclude the electrostatic effect, the used phase hysteresis loops, $\phi(E)$ and amplitude loops $A(E)$ here were obtained at bias-off state. In all the measurements, the scan rate was 1 Hz and the bottom electrode of the sample was grounded (Fig. 1a). A commercial Pt-coated Si tip (AC240TM, Olympus, Japan) with a radius of 15 nm was used in these measurements. The average spring constant is about 2 N/m, and the average resonance frequency is about 65 kHz.

**Electronic transport measurements.** All the electronic transport measurements (sheet resistance, carrier density and mobility) were performed in a Van der Pauw geometry with a Quantum Design Physical Property Measurement System (PPMS). The contacts of the samples were made with an aluminum wire-bonder.

**Scanning transmission electron microscopy.** Scanning transmission electron microscopy (STEM) and electron energy loss spectroscopy (EELS) studies were conducted using a JEOL ARM200F atomic resolution analytical electron microscope equipped with a cold field-emission gun and a new ASCOR 5th order aberration corrector and Gatan Quantum ER spectrometer.

**Data availability**

The data that support the findings of this study are available from the corresponding author on request.

**Acknowledgments**

This work is supported by the National University of Singapore (NUS) Academic Research Fund (AcRF Tier 1 Grants No. R-144-000-364-112 and No. R-144-000-390-114) and the Singapore National Research Foundation (NRF) under the Competitive Research Programs (CRP Awards No. NRF-CRP15-2015-01 and No. NRF-CRP10-2012-02). C. J. Li acknowledges the financial support from Singapore Ministry of Education Academic Research Fund Tier 1 (R-284-000-158-114). S. J. P. thanks Singapore Ministry of Education under its Tier 2 Grant (MOE2017-T2-1-129).

**Author contributions**

W.X.Z. and A.A. conceived and designed the project. W.X.Z. and S.W.Z. prepared the samples and performed the electronic transport measurements. H.J.W., Y.Z., C.J.L. and M.S.L. performed the scanning transmission electron microscopy measurements and analysis under the supervision of S.J.P.. J.Z. performed the first-principles calculations under the supervision of Y.P.F.. J.X.X., R.G. and H.W. performed the piezoresponse force microscopy measurements and analysis under the supervision of K.Y.Z. and J.S.C.. C.G.L. performed the photoluminescence measurements and analysis under the supervision of S.J.C.. P.Y. and W.X.Z. performed the X-ray diffraction measurements and analysis. C.J.L., S.W.Z., Z.H.,

**Figures and Tables**

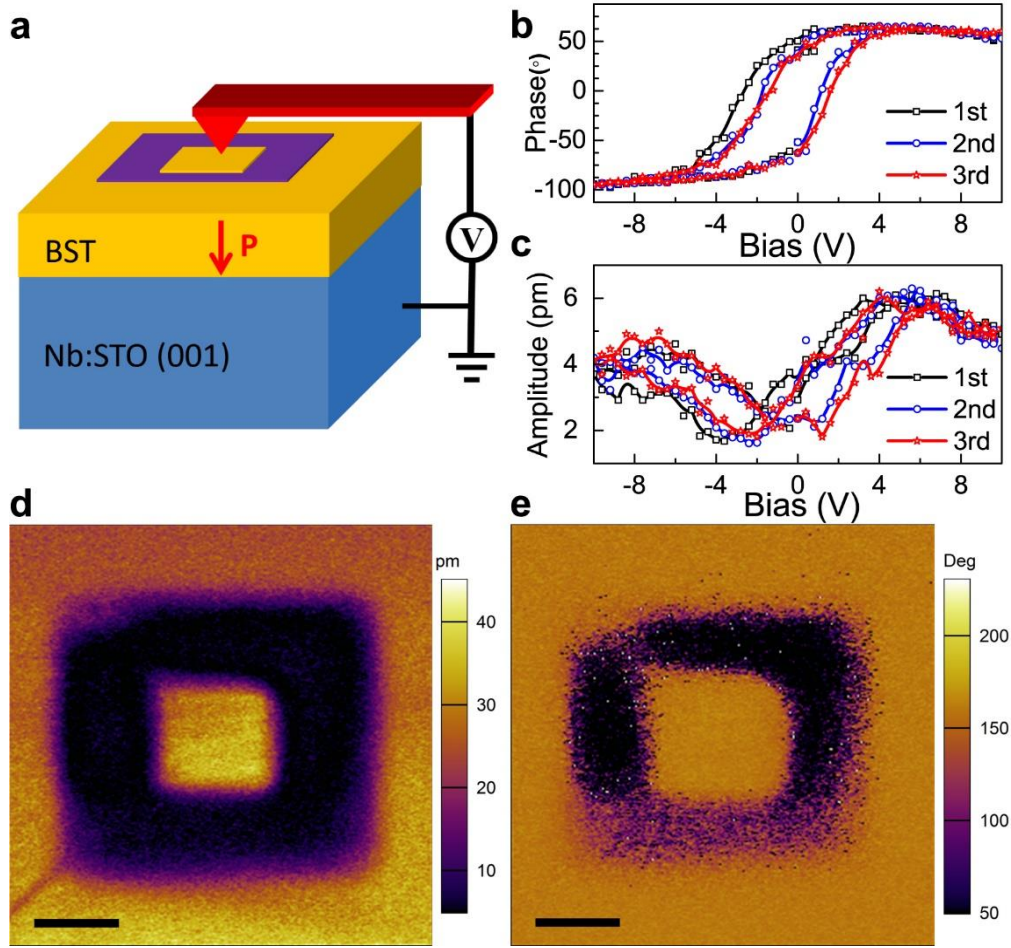

**Fig. 1** Piezoresponse force microscopy characterization of a 10 unit cells of $Ba_{0.2}Sr_{0.8}TiO_3$ thin film at room temperature. **a** Schematic diagram of the geometry for the piezoresponse force microscopy (PFM) setup, the red arrow indicates the as-grown polarization direction of $Ba_{0.2}Sr_{0.8}TiO_3$ (BST). The Nb-doped $SrTiO_3$ (Nb:STO, 0.1 wt% Nb doping) substrate is used as the bottom electrode. **b** Local PFM phase hysteresis loops and, **c** butterfly-like amplitude loops of the BST/Nb:STO sample, different colours stand for different voltage scans, the solid lines are guides for the eyes. **d,e** PFM amplitude (**d**) and phase (**e**) images (5×5 µm$^2$) of the piezoelectric domains after application of -8 and 8 V tip bias to the 3×3 and 1×1 µm$^2$ regions, respectively. Yellow region corresponds to as-grown downward polarization，purple region represents switched upward polarization. The same colour between the as-grown 5×5 µm$^2$

region and the central 1×1 µm$^2$ region indicates that the as-grown BST has a downward polarization. The scale bars in **d** and **e** are 1 µm.

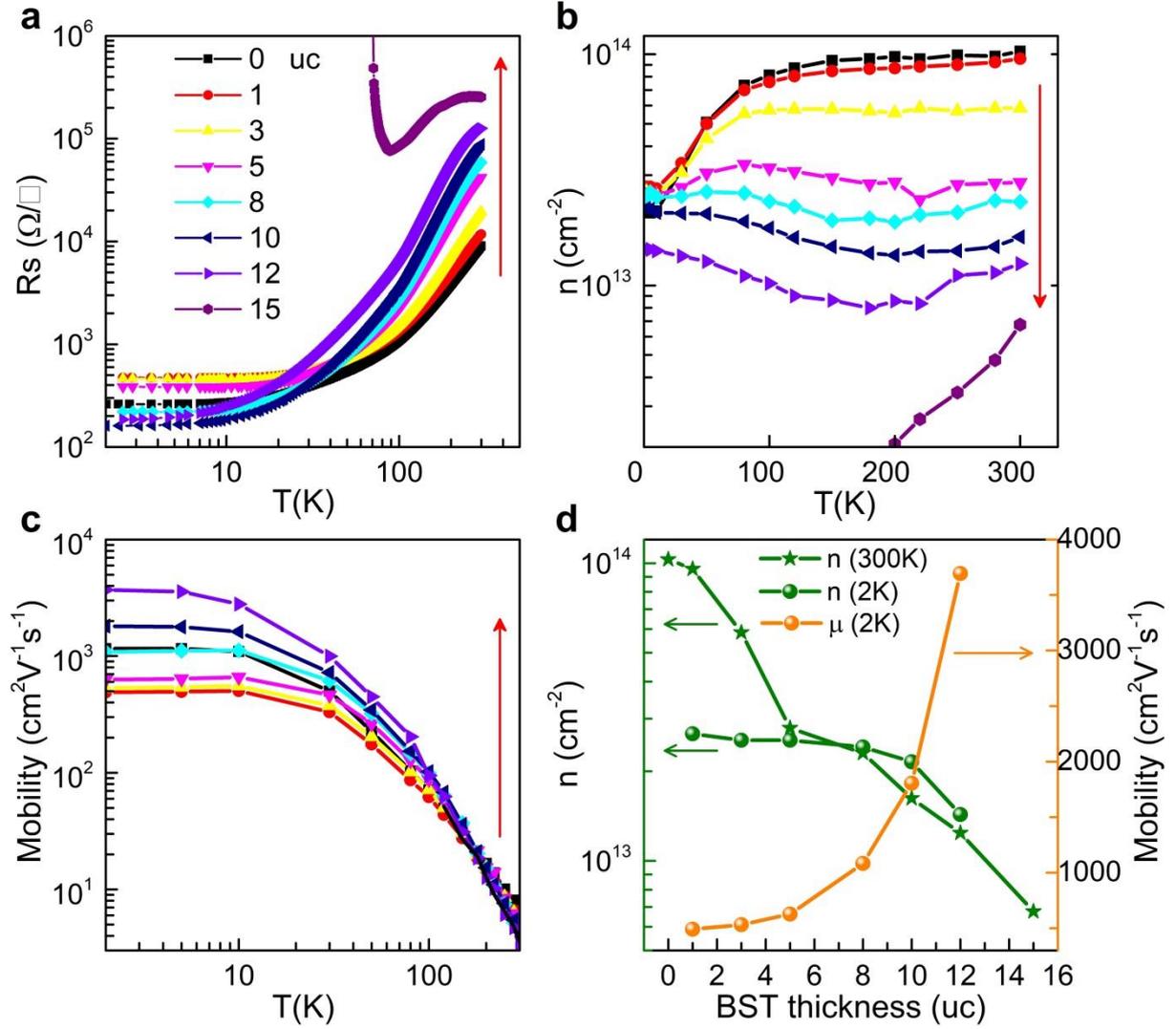

**Fig. 2** Electronic transport characterization of the two-dimensional electron gas in LaAlO$_3$/Ba$_{0.2}$Sr$_{0.8}$TiO$_3$/SrTiO$_3$ heterostructure. **a-c** Temperature dependent sheet resistance $R_s$ (**a**), carrier density $n$ (**b**) and mobility $\mu$ (**c**) for samples with fixed LaAlO$_3$ thickness (15 unit cells) and different unit cells (uc) of Ba$_{0.2}$Sr$_{0.8}$TiO$_3$ (BST) on SrTiO$_3$ substrates. Different colors represent different BST thicknesses. The red arrows indicate increasing BST thickness. **d** A summary of the carrier density (at 300 and 2 K) and mobility (at 2 K) dependence on BST thickness from **b** and **c**.

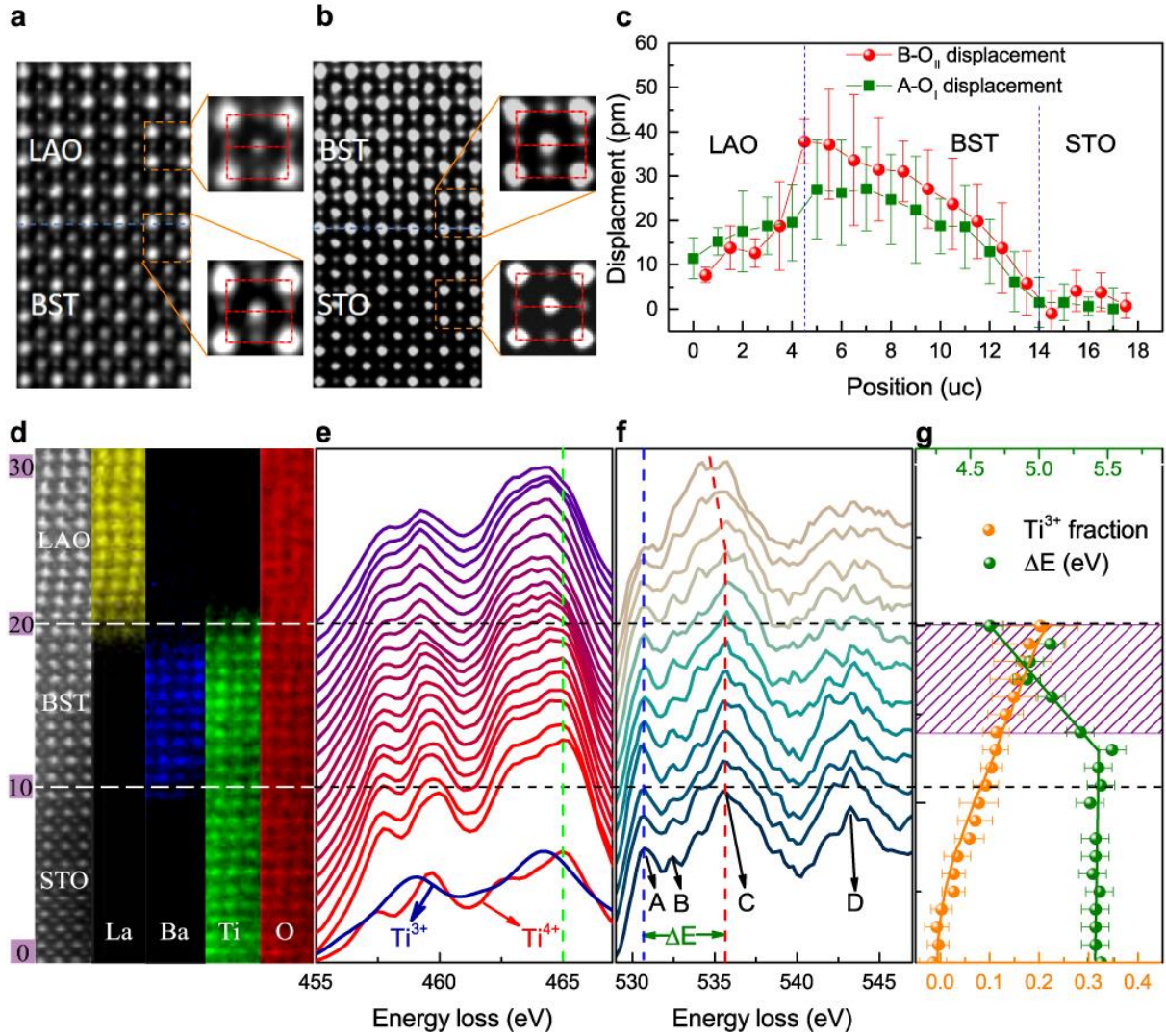

Fig. 3 Structural, elemental and electronic characterization of a LaAlO$_3$/Ba$_{0.2}$Sr$_{0.8}$TiO$_3$/SrTiO$_3$ heterostructure. **a,b** The heterostructure has 15 unit cells (uc) of LaAlO$_3$ (LAO) and 10 unit cells of Ba$_{0.2}$Sr$_{0.8}$TiO$_3$ (BST) on SrTiO$_3$ (STO) substrate. Atomically resolved inverted annular-bright-field scanning transmission electron microscopy (ABF-STEM) images of the LAO/BST interface (a) and BST/STO interface (b), respectively. The orange dashed squares focus on regions which are representative unit cells of LAO, the LAO/BST interface, the BST/STO interface and STO, respectively, and are enlarged for clearer view. The brighter atomic columns in the corner are A-site (Sr, Ba, La) columns and the darker columns at/near the center are B-site (Ti, Al) columns. **c** Out-of-plane B-O$_{II}$ and A-O$_{I}$ displacements across the LAO/BST/STO heterostructure. **d** High-angle annular dark-field

(HAADF) STEM image and electron energy loss spectroscopy (EELS) spectrum images of La $M_{4,5}$ (yellow), Ba $M_{4,5}$ (blue), Ti $L_{2,3}$ (green) and O K (red) edges. $BO_2$ layers are numbered from bottom to top. **e** Layer-resolved Ti $L_{2,3}$ spectra. Reference spectra for $Ti^{4+}$ (red) and $Ti^{3+}$ (blue) are shown at the bottom, taken from $SrTiO_3$ and $Ti_2O_3/Al_2O_3$, respectively. **f** Layer-resolved O K spectra, which are normalized by their respective main peaks near 535 eV. **g** Layer-resolved $Ti^{3+}$ fraction (orange), which is defined as $Ti^{3+}/(Ti^{3+}+Ti^{4+})$, and $\mathit{\Delta E}$ from the O K edge (olive). The purple shaded area indicates the approximate location of the 2DEG. **e-g** are all aligned with the HAADF image and EELS spectra in **d**. The error bars show the standard deviations of the averaged measurements for each vertical atomic layer.

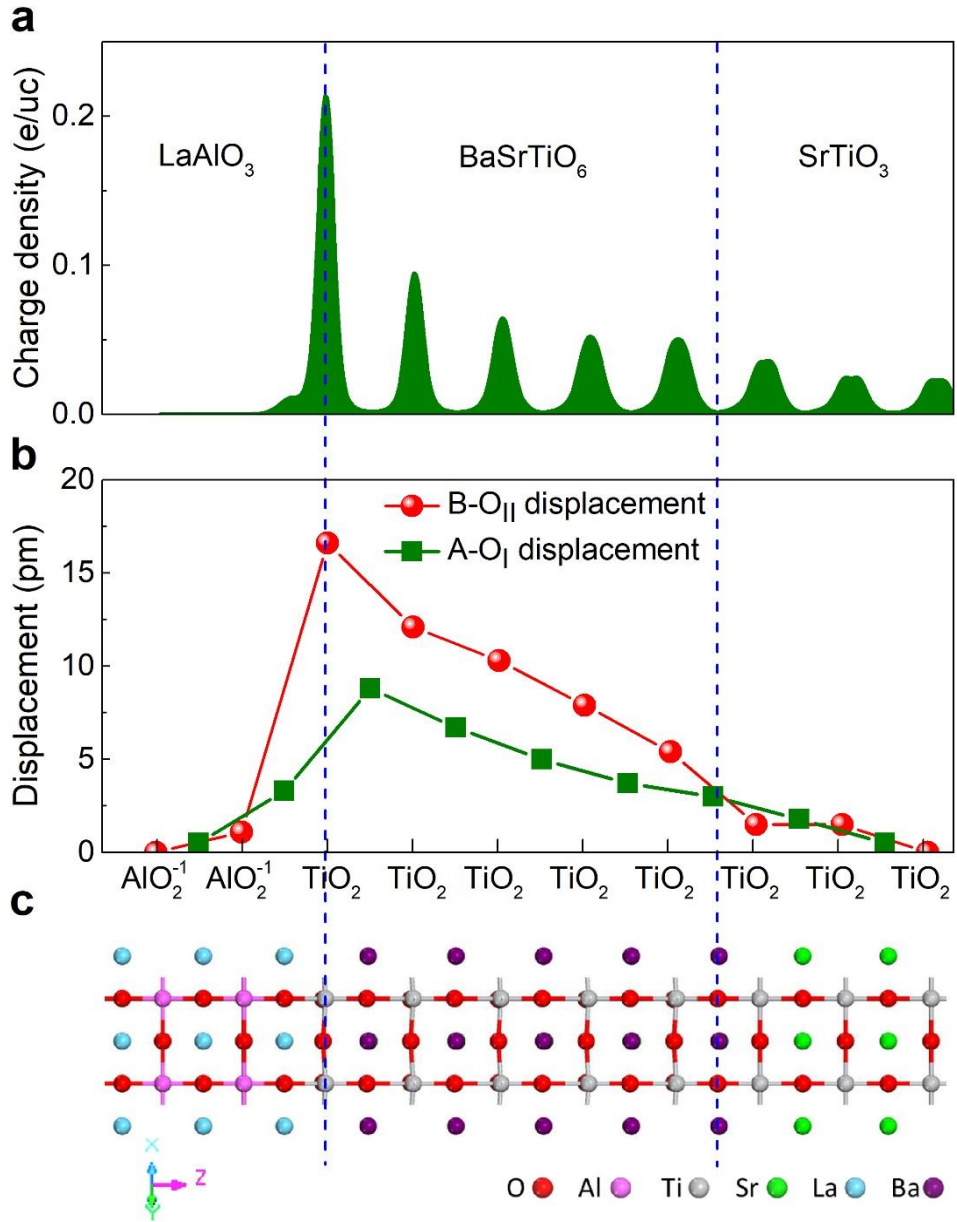

**Fig. 4** First-principles calculations of the two-dimensional electron gas and polar displacements in LaAlO$_3$/Ba$_{0.5}$Sr$_{0.5}$TiO$_3$/SrTiO$_3$ heterostructure. **a** Layer-resolved in-plane average charge density along the <001> direction. **b** Layer-resolved out-of-plane displacements of B-O$_{II}$ (red balls) and A-O$_I$ (green squares), respectively. **c** Structural guide of the supercell used in the calculations (only half of the structure used in the calculations is shown).

**Supplementary note 1**

**Amorphous LaAlO$_3$ on BaTiO$_3$ (001) substrate.** We have deposited amorphous LaAlO3 (a-LAO) on BaTiO$_3$ (BTO) substrate in an attempt to induce oxygen vacancy in BTO and thus create a two-dimensional electron gas (2DEG) at the a-LAO/BTO interface. This method is very similar to the method used by Rödel *et al.* in principle[1]. As shown in supplementary Fig. 1, we deposited 10 nm of a-LAO on BTO substrate and the sample shows a semiconducting behavior and strong localization at low temperatures (supplementary Fig. 1a). At 300 K, the 2DEG has a carrier density $n_{2D} \approx 8 \times 10^{13}$ cm$^{-2}$ with a mobility of 0.6 cm$^2$V$^{-1}$s$^{-1}$ (supplementary Fig. 1b). On the other side, a LaAlO$_3$/SrTiO$_3$ (LAO/STO) sample prepared under similar conditions has $n_{2D} \approx 1 \times 10^{14}$ cm$^{-2}$ with a mobility of 5-6 cm$^2$V$^{-1}$s$^{-1}$ (see main text Fig. 3, the 15 unit cells LAO/STO sample). Hence, one can clearly see that the 2DEG density in BTO is not only smaller as compared to that observed with angle-resolved photoemission spectroscopy (ARPES) in supplementary Reference 1, but also smaller than that in LAO/STO. Furthermore, the mobility of the 2DEG in BTO is one order of magnitude smaller than that of LAO/STO, indicating strong localization of electrons in BTO due to strong Coulomb forces imposed by ferroelectricity. In addition, the a-LAO/BTO samples show very non-uniform conductivity, which could be because that the ferroelectric ordering and metallic conduction occur in two distinct regions. These results suggest at BTO is not an ideal candidate to study the coexistence of ferroelectricity and 2DEG.

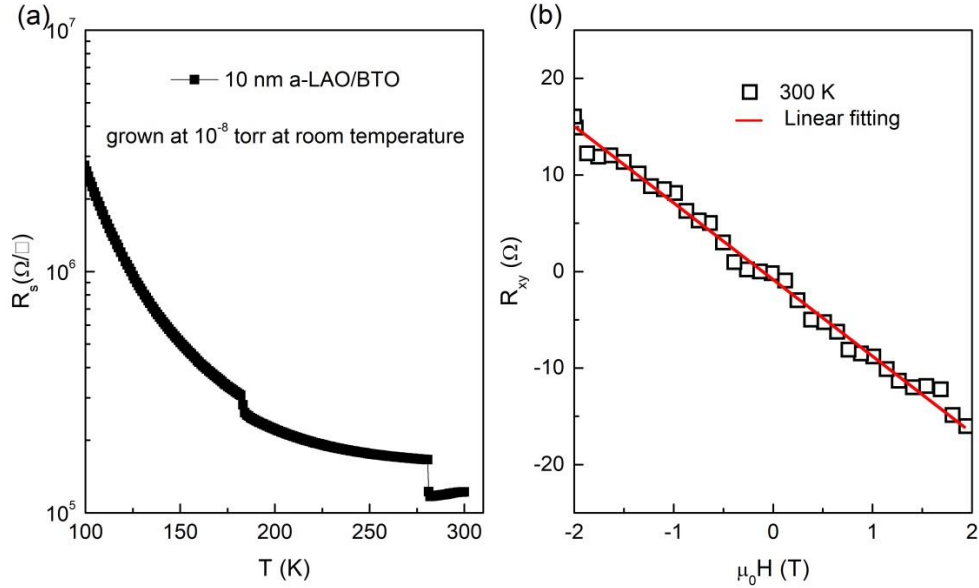

**Supplementary Figure 1. Strong localization of electrons at a-LAO/BTO interface.** Sheet resistance dependence on temperature (**a**) and Hall effect at 300 K (**b**) for a 10 nm a-LAO/BTO sample grown at $10^{-8}$ torr at room temperature. The red line in (**b**) is a linear fitting of the Hall effect.

**Supplementary note 2**

**Epitaxial growth of $Ba_{0.2}Sr_{0.8}TiO_3$ (BST) and LAO on STO (001).** All the LAO/BST/STO (001) heterostructures were grown by pulsed laser deposition (PLD) equipped with *in-situ* reflection high-energy electron diffraction (RHEED). Supplementary Fig. 2a-c show RHEED patterns for STO (001) substrate, after BST deposition and after LAO deposition, respectively. The clear Laue circles and Kikuchi lines indicate two-dimensional growth of both BST and LAO. The RHEED oscillations for a 15 uc LAO/25 uc BST/STO sample indicates layer-by-layer growth of both BST and LAO, as shown in Supplementary Fig. 2d.

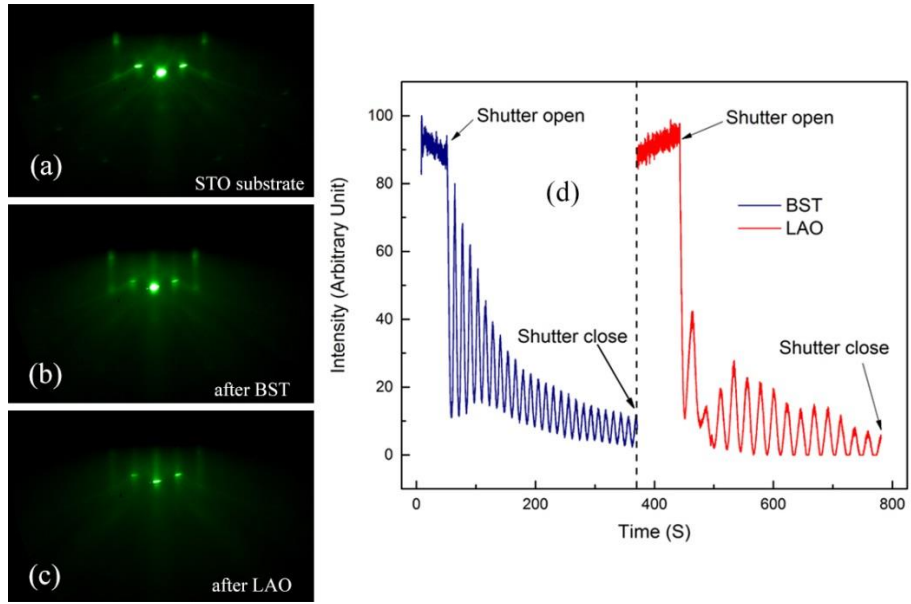

**Supplementary Figure 2. Epitaxial growth of BST and LAO on STO (001).** (**a-c**) RHEED patterns for STO (001) substrate, after BST deposition and after LAO deposition, respectively. (**d**) RHEED oscillations for a 15 uc LAO/25 uc BST/STO sample.

**Supplementary note 3**

**Surface characterization of LAO/BST/STO samples. Supplementary** Figure 3 shows atomic force microscopy (AFM) image of a 15 uc LAO/10 uc BST/STO sample. The AFM image indicates that the surface of the film is atomically smooth.

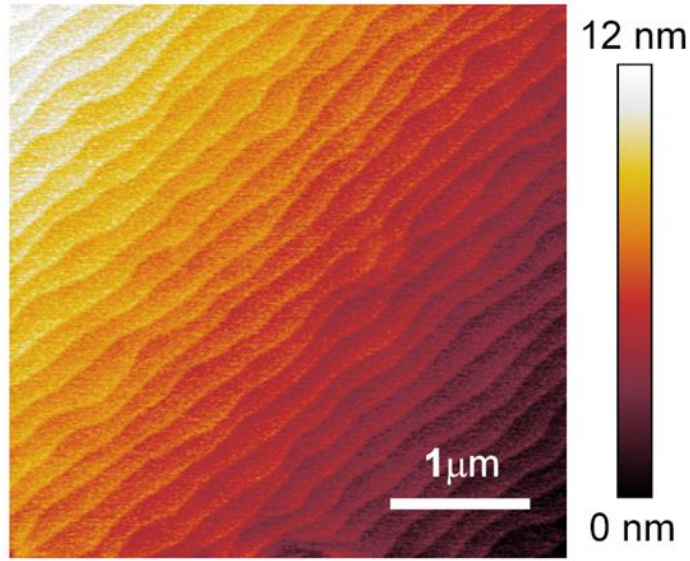

**Supplementary Figure 3. Surface characterization of LAO/BST/STO heterostructure.** Atomic force microscopy image of a 15uc LAO/10uc BST/STO sample.

**Supplementary note 4**

**Crystalline structure determination of BST thin films.** To find out the strain state of the BST thin films, we performed X-ray diffraction (XRD) reciprocal space mapping (RSM) for a 10 nm BST/STO sample. Supplementary Fig. 4 shows the RSM around (002) (a,b), (-103) (c) and (0-13) (d) peaks. One can see that BST is coherently grown on STO substrate without relaxation, i.e. the in-plane lattice constant of BST is fully constrained to that of STO substrate. From these results we calculated the lattice parameters of BST thin film as shown in supplementary Table 1. 10 nm BST grown on STO shows a tetragonal structure with $a=b=3.9051$ Å, $c=3.9713$, $\alpha=\beta=\gamma=90°$. So that $c/a$ ratio is 1.017, which is larger than the $c/a$ ratio of bulk $BaTiO_3$ (1.011). This large $c/a$ ratio is probably responsible for room-temperature ferroelectricity in our BST thin films.

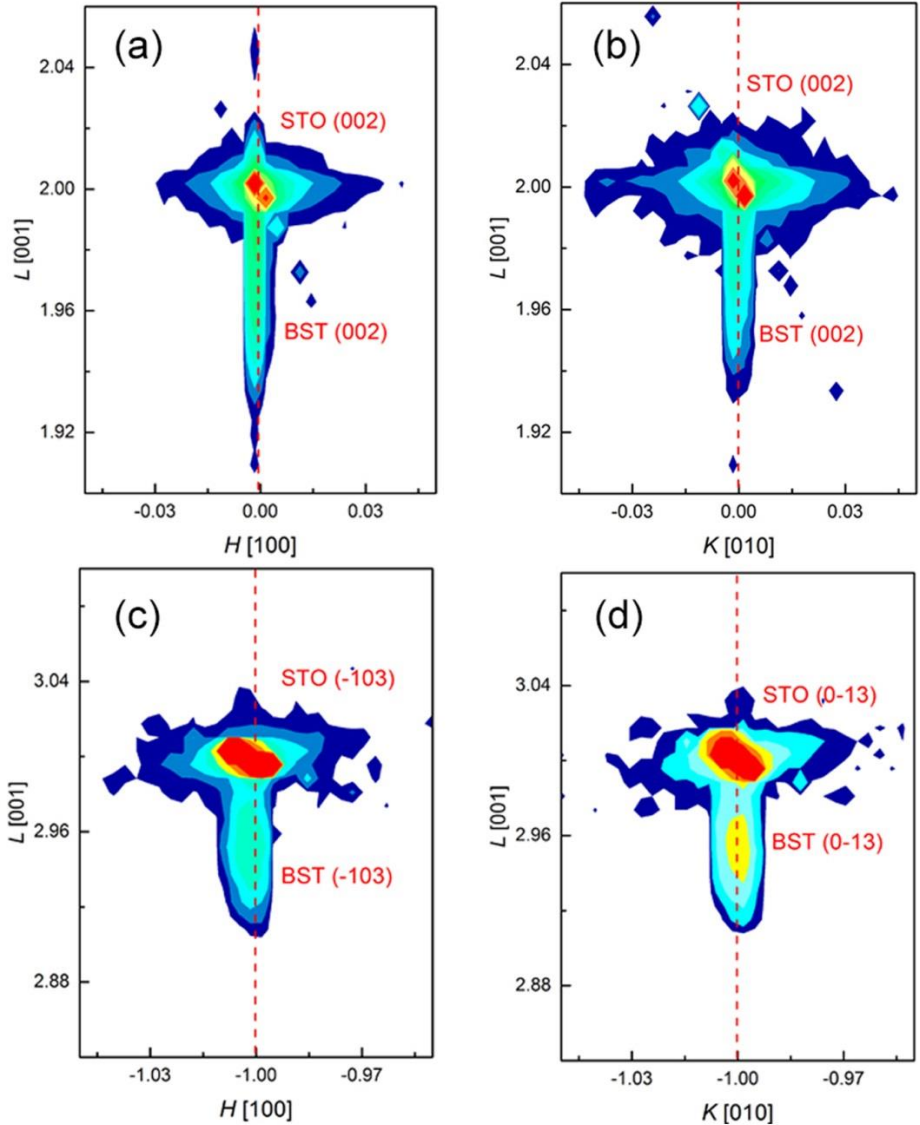

**Supplementary Figure 4. Crystalline structure determination of BST thin films**. XRD reciprocal space mapping around (002) (**a,b**), (-103) (**c**) and (0-13) (**d**) peaks of a 10 nm BST/STO sample. The red spots in the figures represent the substrate peak, while the film peak with a much lower intensity lies below the substrate peak.

|  | $a$ (Å) | $b$ (Å) | $c$ (Å) | $\alpha$ (°) | $\beta$ (°) | $\gamma$ (°) |
|---|---|---|---|---|---|---|
| Value | 3.9051±0.001 | 3.9051±0.001 | 3.9713±0.001 | 90.0000±0.1714 | 90.0000±0.1038 | 90.0000±0.1377 |

**Supplementary Table 1. Calculated lattice parameters and crystalline angles of 10 nm BST grown on STO (001) substrate.**

**Supplementary note 5**

**Hall effect of the LAO/BST/STO samples.** All sheet carrier concentrations ($n$) and Hall mobilities ($\mu$) were calculated according to $n = -1/(eR_{xy})$ and $\mu = (enR_S)^{-1}$, where e is the electron charge, $R_s$ is sheet resistance and $R_{xy}$ is Hall resistance. Supplementary Fig. 5 clearly shows that the carrier density decreases with increasing BST thickness as the slope of the Hall effect becomes steeper (supplementary Fig. 5a). In addition, by comparing the temperature dependence of the Hall effect between a 15 uc LAO/1 uc BST/STO (supplementary Fig. 5b) and a 15 uc LAO/10 uc BST/STO sample (supplementary Fig. 5c), one can clearly see that the carrier density of the 15 uc LAO/1 uc BST/STO sample is strongly temperature dependent, while the carrier density of the 15 uc LAO/10 uc BST/STO sample is nearly temperature independent.

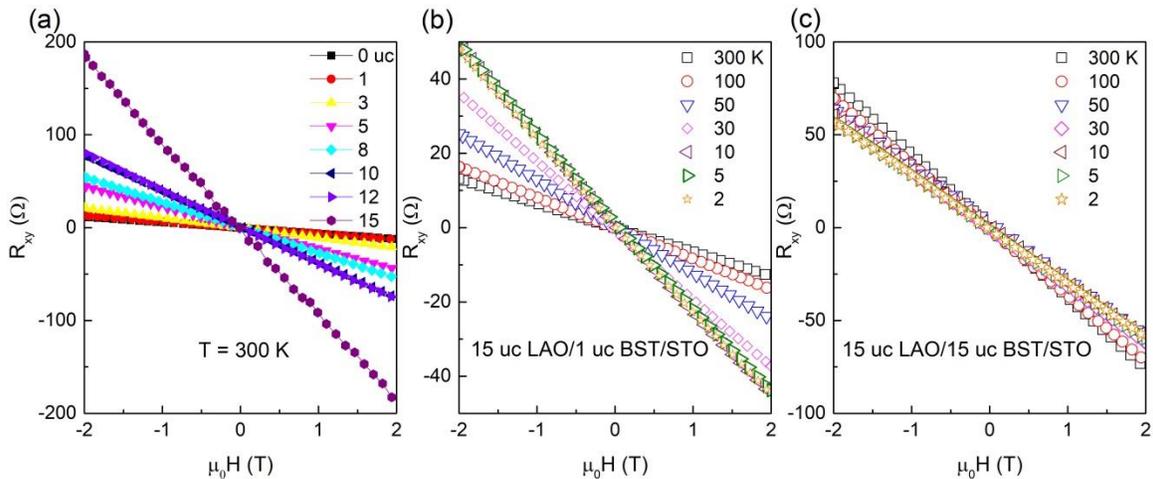

**Supplementary Figure 5. Hall effect of the LAO/BST/STO samples.** Hall effect for LAO/BST/STO samples with 15 uc of LAO and different thicknesses of BST at 300 K (**a**).

Hall effect for a 15 uc LAO/1 uc BST/STO (**b**) and a 15 uc LAO/10 uc BST/STO sample (**c**) at different temperatures.

**Supplementary note 6**

**LAO/BST/STO samples with fixed BST thickness and different LAO thicknesses.** To verify the competition between the built-in electric field of LAO and the polarization of BST, another set of sample with fixed BST thickness (10 uc) and different LAO thicknesses (0-25 uc) was prepared. Supplementary Fig. 6a shows temperature dependent sheet resistance ($R_s$-$T$) for these samples. When LAO thickness $t$ is below 8 uc, the samples show bad conductivity and the data are not shown here. For 8 uc $\leq t <$ 20 uc, $R_s$-$T$ curves show fully metallic behavior from 300 K to 2 K. When $t \geq$ 20 uc, an upturn at around 40 K is observed, which is probably due to Kondo effect as also observed in LAO/STO when LAO thickness is larger than 26 uc[2] or increasing interfacial disorder with increasing film thickness and strain effects introduced by substrate misfit as discussed in the main text[3]. Due to the electric field produced by BST, the critical thickness of LAO for insulator-metal transition should be larger than that of LAO/STO heterostructure. Experimentally, the LAO critical thickness was found to be 5-8 uc, which is larger than that of LAO/STO (4 uc). As shown in supplementary Fig. 6b, the conductance of the 2DEG increases by five orders from 5 to 8 uc of LAO. Here, conductance is defined as $\sigma$=1/$R_s$. Supplementary Fig. 6c,d show carrier density and mobility dependence on temperature for LAO/BST/STO samples with fixed BST thickness (10 uc) and different LAO thickness, respectively. Carrier density and mobility of LAO/STO samples are also shown to give a comparison. As can be seen in supplementary Fig. 6c, all LAO/BST/STO samples show smaller carrier density than that of LAO/STO samples. And all LAO/BST/STO samples show temperature independent carrier density, as opposed to LAO/STO samples, whose carrier density decreases as temperature decreases. These features indicate ferroelectric depletion of carriers and suppression of oxygen vacancy formation in LAO/BST/STO samples as explained in the main text. As shown in supplementary Fig. 6d, all LAO/BST/STO samples show higher mobility (1500-2000 $cm^2V^{-1}s^{-1}$) than that of LAO/STO samples (500-1000 $cm^2V^{-1}s^{-1}$) at 2 K. This is due to less electron scattering and defect scattering as explained in main text.

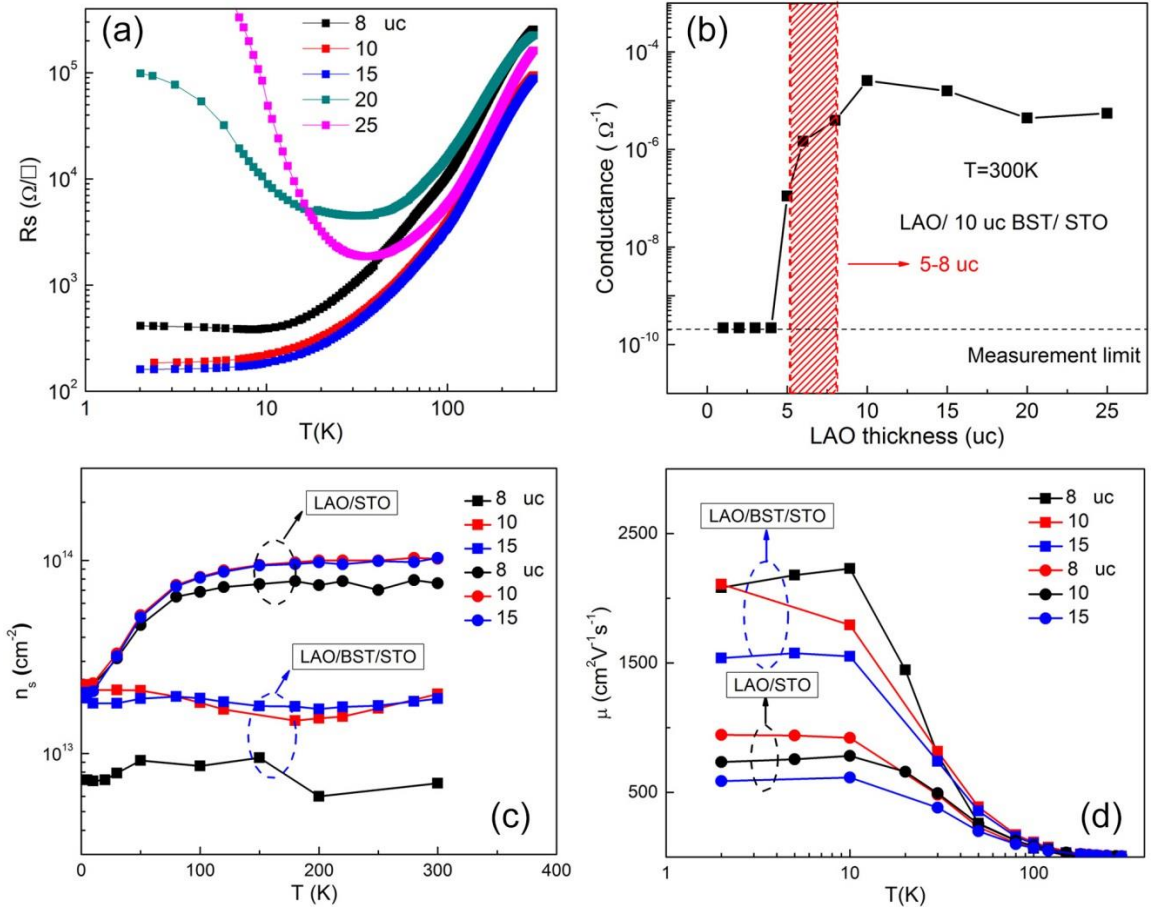

**Supplementary Figure 6. Electronic transport characterization of the 2DEG in LAO/BST/STO heterostructure with fixed BST thickness (10 uc) and different LAO thicknesses.** (**a**) Temperature dependent sheet resistance ($R_s$) for samples with 8, 10, 15, 20, 25 uc of LAO. (**b**) Conductance dependence on LAO thickness for the samples with 10

uc BST and different LAO thicknesses. (**c**) Temperature dependent carrier density ($n$) and (**d**) mobility ($\mu$) for LAO/BST/STO samples with fixed BST thickness (10 uc) and different LAO thicknesses compared to that of LAO/STO samples with different LAO thicknesses. Different colors represent different LAO thicknesses, the solid squares stand for LAO/BST/STO samples and the solid circles indicate LAO/STO samples.

**Supplementary note 7**

**Photoluminescence measurement of oxygen vacancy concentrations.** To detect the oxygen vacancy concentrations, we carried out photoluminescence (PL) measurements on the samples with fixed LAO thickness (15 uc) and different BST thicknesses. Room temperature PL spectra were recorded using the 325 nm line of a He-Cd laser as the excitation source. The laser power is about 105 mW. As shown in supplementary Fig. S7, the peak position (490 nm) agrees well with the wavelength reported in supplementary Reference 4. As discussed in the main text, the PL intensity is proportional to oxygen vacancy concentration in STO[4,5]. One can see that the PL intensity gradually decreases with increasing BST thickness, indicating less oxygen vacancies. When BST exceeds 8 uc, the samples show comparable PL intensity to that of an as-received STO substrate, indicating that they are (nearly) free from oxygen vacancies. When BST thickness is below 8 uc (~3.2 nm), 2DEG resides both in STO and BST, and a large number of oxygen-vacancy-induced carriers are present in STO. When BST thickness exceeds 8 uc, 2DEG resides mainly in BST. Hence, few oxygen-vacancy-induced carriers are present and the carriers mainly come from the polar discontinuity mechanism, leading to observation of temperature-independent carrier density. These results are in good agreement with previous observations, which showed the thickness of the 2DEG at LAO/STO interface is around 2-4 nm[6,7].

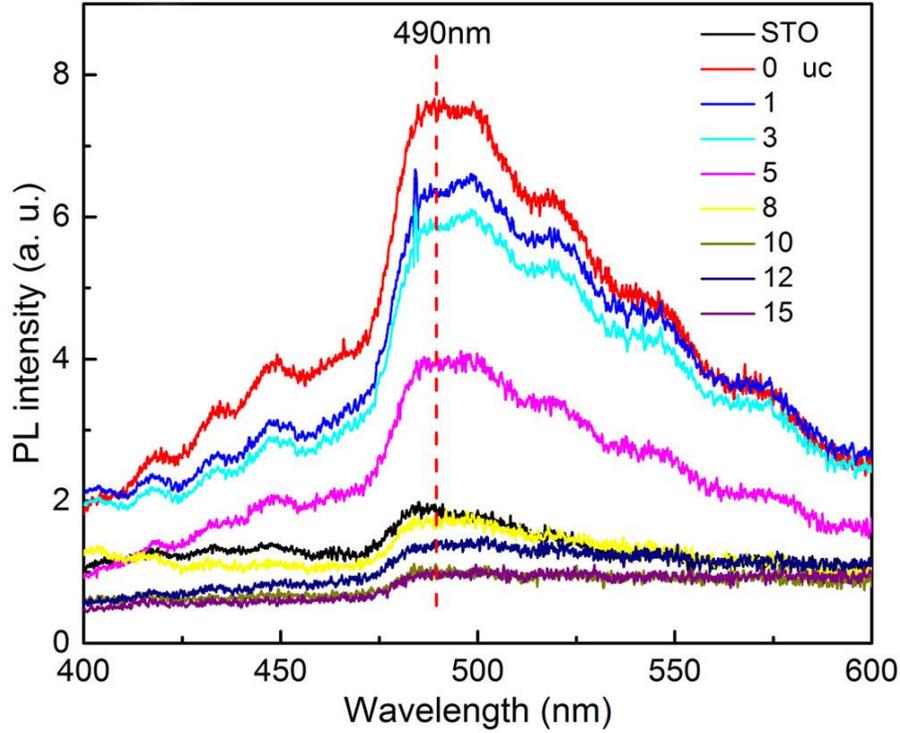

**Supplementary Figure 7. Photoluminescence measurement of oxygen vacancy concentrations**. Room-temperature PL spectra of the same samples in Fig. 2 of the main text and an as-received STO substrate. Here, STO substrate is used as a reference sample, which is (nearly) free of oxygen vacancy. Different colors indicate different thicknesses of BST, while LAO is fixed to 15 uc.

**Supplementary note 8 Scattering mechanism analysis of the 15 uc LAO/15 uc BST/STO sample.** The upturn of the $R_s$-$T$ curve and the corresponding decreasing carrier density ($n$) with decreasing temperature of the 15 uc BST sample (Fig. 2a,b) suggest carrier localization and is consistent with Mott variable-range hopping (VRH) model. Supplementary Fig. 8 shows the fitting of the $R_s$-$T$ curve of a 15 uc LAO/15 uc BST/STO with the VRH model, where $R_s = R_0\exp[(T_0/T)^{1/3}]$[3].

As explained in the main text, carrier density gradually decreases with increasing BST thickness as the magnitude of the polarization increases (Fig. 2b,d)[8], which obstructs the charge transfer from LAO surface to LAO/BST interface. When BST is 15 uc, the polarization becomes strong enough to deplete most of the mobile electrons and the remaining electrons lie below the mobility edge. These remaining electrons can be excited by thermal energy at high temperature and become localized at low temperature. This explains the upturn of the $R_s$-$T$ curve and the corresponding decreasing carrier density ($n$) with decreasing temperature.

We also note, that the electron localization of epitaxial LaAlO$_3$/SrTiO$_3$ interface has been studied in various systems such as LaAlO$_3$/SrTiO$_3$/NdGaO$_3$ (LAO/STO/NGO)[9] and LaAlO$_3$/SrTiO$_3$/ LaAlO$_3$)$_{0.3}$-(Sr$_2$AlTaO$_3$)$_{0.7}$ (LAO/STO/LSAT)[3]. In these literatures, several other mechanisms have been proposed, such as increasing interfacial disorder with increasing film thickness, strain effects introduced by substrate misfit and larger electron mass of the electrons in $d_{xz}$ and $d_{yz}$ bands. We believe these mechanisms could also collectively contribute to the electron localization in the 15 uc LAO/15 uc BST/STO sample.

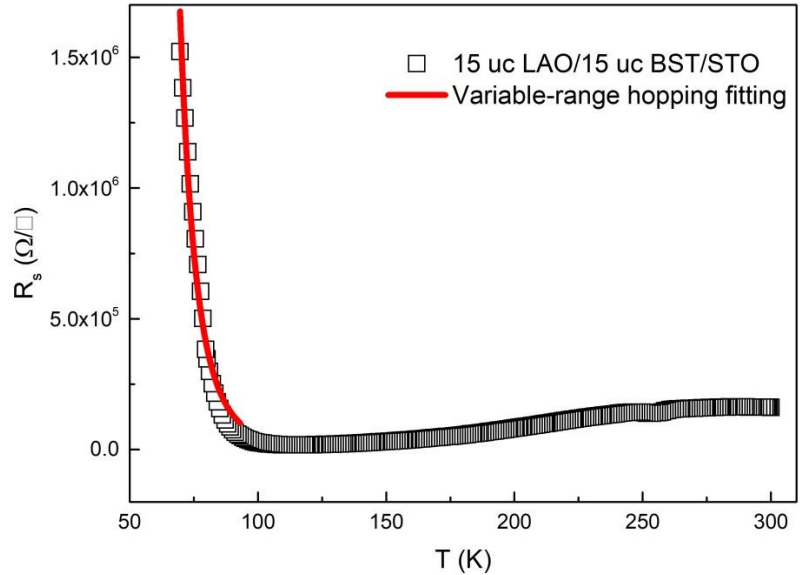

**Supplementary Figure 8. Scattering mechanism analysis of the 15 uc LAO/15 uc BST/STO sample.** Sheet resistance dependence on temperature for a 15 uc LAO/15 uc BST/STO sample (black hallow squares) with variable-range hopping fitting at low temperatures (red solid line).

**Supplementary note 9**

**High-angle annular dark-field and annular bright-field scanning transmission electron microscopy images.** The overall view of the LAO/BST/STO heterostructure can be seen in low-magnification scanning transmission electron microscopy (STEM) high-angle annular dark-field (HAADF) image (supplementary Fig. 9a), and both LAO/BST and BST/STO interfaces are sharp and fully strained. Supplementary Fig. 9b,c show a pair of STEM HAADF and annular bright-field (ABF) images of LAO/BST/STO. STEM HAADF and ABF have been proved to be an effective structure imaging mode, producing contrast interpretable by $Z$ contrast (the signal is roughly proportional to $Z^2$ for HAADF and $Z^{1/3}$ for ABF)[10,11]. Therefore, the regions of LAO, BST films and STO substrate as well as their interfaces can be further differentiated, as shown from the integrated intensity profile inserted in supplementary Fig. 9b. The lattices between the three components are well aligned along the in-plane direction, while exhibiting a small difference along the out-of-plane direction, as shown in the fast Fourier transform (FFT) image (inset in supplementary Fig. 9c). To determine the polarization vectors, especially around the LAO/BST interface, the atom positions of supplementary Fig. 9b,c were located accurately by fitting them as 2D Gaussian peaks[12]. The polarization points from the centre of the four corner A-site atoms to B-site atom.

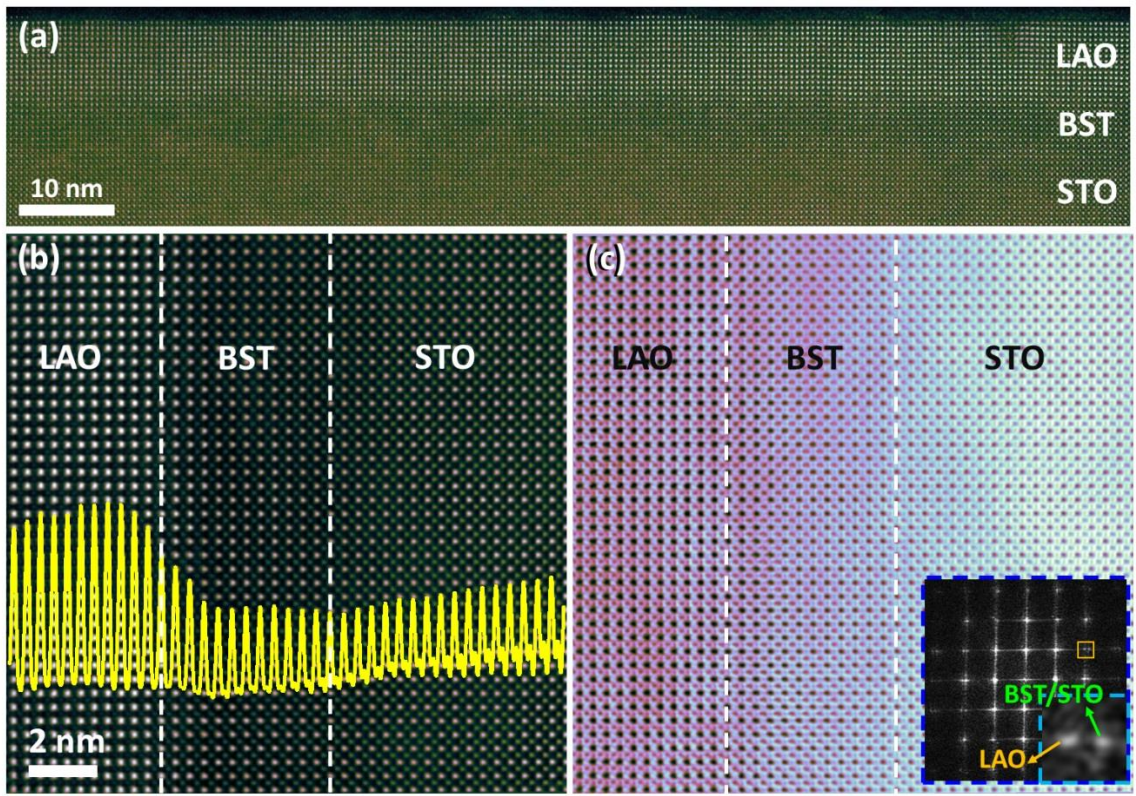

**Supplementary Figure 9. STEM images of a 15 uc LAO/10 uc BST/STO sample.** (**a**)Large-view HAADF-STEM image (colored). (**b,c**) Atomically resolved HAADF-STEM (**b**) and ABF-STEM (**c**) images (colored), the inset in (**b**) is an integrated intensity profile; the inset in (**c**) is a FFT image**,** where split peaks are marked.

**Supplementary note 10**

**Numerical fitting of Ti L$_{2,3}$ spectra at LAO/BST interface.** To trace the Ti valence change quantitatively across the LAO/BST interface, we performed multiple linear least squares (MLLS) fitting of the experimental spectra. Supplementary Fig. 10a focuses on BST and LAO/BST interface regions. The reference Ti$^{4+}$ and Ti$^{3+}$ spectra were obtained from the STO region well away from the interface and Ti$_2$O$_3$/Al$_2$O$_3$ heterostructure with stable Ti$^{3+}$ valence respectively, as shown in supplementary Fig. 10b. The experimental Ti L$_{2,3}$ edge is similar with that acquired from LaTiO$_3$/SrTiO$_3$ with partially oxidized Ti, which hints that the Ti valence around the LAO/BST interface is lower than 4+. Supplementary Fig. 10c,d are MLLS fitted spectrum images (colorized) of Ti$^{4+}$ and Ti$^{3+}$ edges of BST and LAO/BST interface regions, respectively. It is clearly seen that the intensity of Ti$^{4+}$ gradually decreases (color changes to green and blue) when approaching LAO/BST interface from BST, while Ti$^{3+}$ signal increases (color changes to red and yellow) correspondingly. The quantitative analysis of Ti$^{3+}$ fraction is summarized in Fig. 3g of the main text.

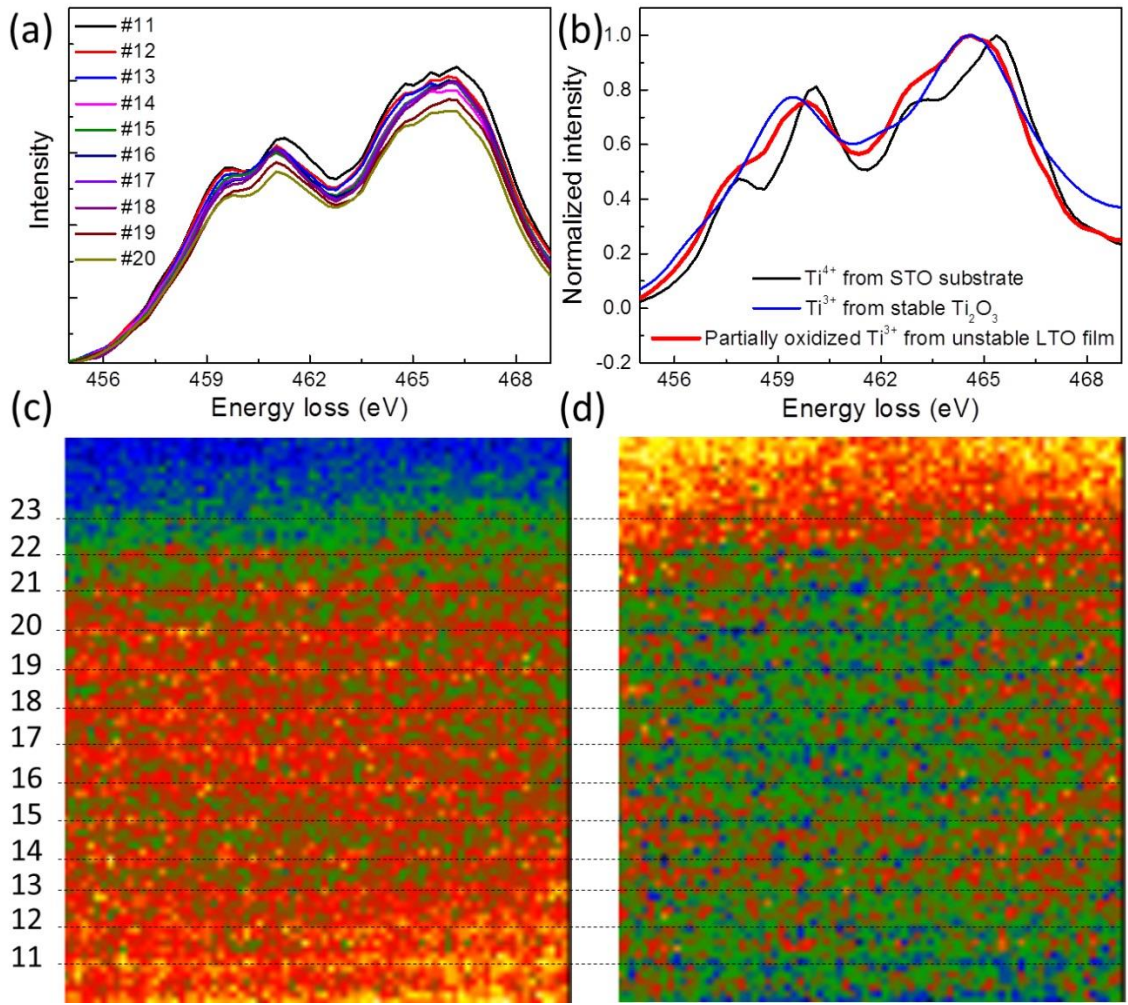

**Supplementary Figure 10. Numerical fitting of Ti L$_{2,3}$ spectra at LAO/BST interface.** (**a**) Layer-by-layer Ti L$_{2,3}$ spectra focusing on BST region. (**b**) Comparison between Ti L$_{2,3}$ edge in STO well away from the interface as the Ti$^{4+}$ reference spectrum (black line), Ti L$_{2,3}$ edge from Ti$_2$O$_3$/Al$_2$O$_3$ heterostructure with stable Ti$^{3+}$ valence as the Ti$^{3+}$ reference spectrum (blue line), and Ti L$_{2,3}$ edge from LaTiO$_3$/SrTiO$_3$ with partially oxidized Ti$^{3+}$ spectrum (red line), which is similar with the present experimental spectra at the LAO/BST interface. (**c,d**) The MLLS fitted spectrum images (colorized) of Ti$^{4+}$ and Ti$^{3+}$ edges of BST and LAO/BST interface regions (from #11 to #23), respectively.

**Supplementary note 11**

**First-principle simulation of polarization direction of BST on STO substrate.** As shown in supplementary Fig. 111, we found that the spontaneous polarization direction of BST on STO is pointing-down. Supplementary Fig. 11 a,c show side and top view of the simulation cell of the LAO/ $Ba_{0.5}Sr_{0.5}TiO_3$/STO heterostructure. Supplementary Fig. 11b shows that $Ti^{4+}$ and $Sr^{2+}$ (or $Ba^{2+}$) ions are displaced downward relative to $O^{2-}$ with the angel between two $Ti^{4+}$-$O^{2-}$ bonds of 173.557°. This displacement creates a downward polarization in BST.

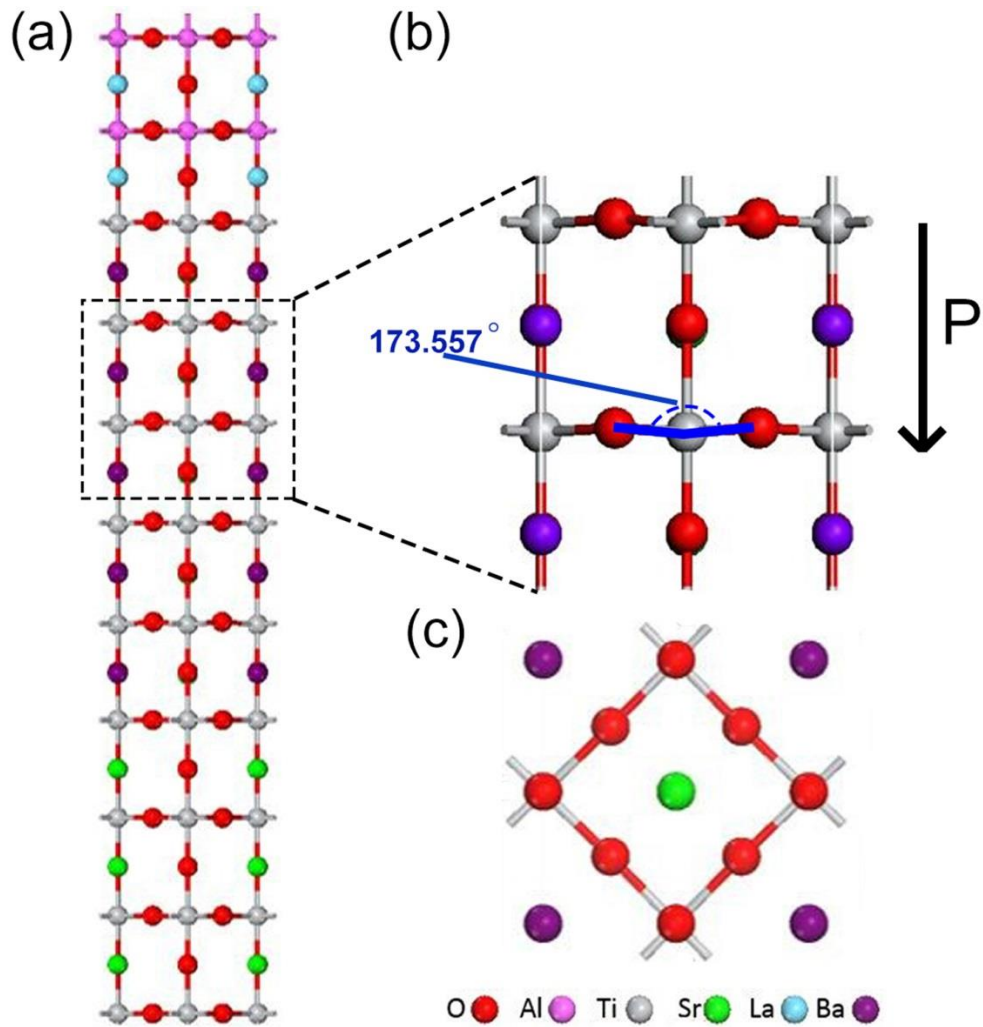

**Supplementary Figure 11. Polarization direction of BST.** (**a**) Side view of LaAlO$_3$/Ba$_{0.5}$Sr$_{0.5}$TiO$_3$/SrTiO$_3$ heterostructure. (**b**) Zoom-in view of Ba$_{0.5}$Sr$_{0.5}$TiO$_3$ layer showing spatial displacement of Ti-O and Sr (or Ba)-O. This displacement creates a downward polarization in Ba$_{0.5}$Sr$_{0.5}$TiO$_3$. (**c**) Top view of the Ba$_{0.5}$Sr$_{0.5}$TiO$_3$ layers.